\documentclass[aps,showpacs,nofootinbib]{revtex4}

\usepackage{graphicx}
\usepackage{amsmath}
\usepackage{amssymb}

\newcommand{\be}{\begin{equation}}
\newcommand{\ee}{\end{equation}}
\newcommand{\ba}{\begin{eqnarray}}
\newcommand{\ea}{\end{eqnarray}}

\def\lb{\label}

\begin{document}

\title{Spontaneous rotational symmetry breaking and 
roton like excitations in gauged $\sigma$-model at finite density}

\author{V.P. Gusynin}
  \email{vgusynin@bitp.kiev.ua}
  \altaffiliation[On leave from ]{
       Bogolyubov Institute for Theoretical Physics,
       03143, Kiev, Ukraine
}

\author{V.A. Miransky}
  \email{vmiransk@uwo.ca}
   \altaffiliation[On leave from ]{
       Bogolyubov Institute for Theoretical Physics,
       03143, Kiev, Ukraine
}
   
\affiliation{
Department of Applied Mathematics, University of Western
Ontario, London, Ontario N6A 5B7, Canada
}

\author{Igor Shovkovy}
  \email{shovkovy@th.physik.uni-frankfurt.de}
  \altaffiliation[On leave from ]{%
       Bogolyubov Institute for Theoretical Physics,
       03143, Kiev, Ukraine
}%
\affiliation{%
       Institut f\"{u}r Theoretische Physik,
       J.W. Goethe-Universit\"{a}t,
       D-60054 Frankurt/Main, Germany
}%

\date{\today}

\begin{abstract}
The linear $\sigma$-model with a chemical potential for hypercharge
is a toy model for the description of the dynamics of the kaon condensate
in high density QCD. We analyze the dynamics of the gauged version 
of this model. It is shown that spontaneous breakdown of 
$SU(2) \times U(1)_{Y}$ symmetry, caused by the chemical potential,
is always accompanied by spontaneous breakdown of both rotational
symmetry and electromagnetic $U(1)_{em}$. The spectrum of excitations
in this model is rich and, because of rotational symmetry breakdown, 
anisotropic. It is shown that there
exist excitation branches that behave as phonon like quasiparticles 
for small momenta and as roton like ones for large momenta. This
suggests that this model can be relevant for anisotropic 
superfluid systems. 
\end{abstract}

\pacs{11.15.Ex, 11.30.Qc}


\maketitle

\section{Introduction}

Recently a class of relativistic models with a finite density of 
matter has been revealed in which spontaneous
breakdown of continuous symmetries leads to a lesser number of
Nambu-Goldstone (NG) bosons than that required by the Goldstone
theorem \cite{MS,STV}. It is noticeable that this class, in
particular, describes the dynamics of the kaon condensate in
the color-flavor locked phase of high density QCD 
that may exist in cores of compact stars
\cite{BS}.

The simplest representative of this class is the linear
$SU(2)_{L}\times SU(2)_R$  $\sigma$-model with the chemical potential
for the hypercharge $Y$:
\ba
{\cal L} = (\partial_{0} +i \mu )\Phi^{\dagger}
(\partial_{0} -i \mu )\Phi 
-\partial_{i}\Phi^{\dagger}
\partial_{i}\Phi -m^{2}\Phi^{\dagger} \Phi
-\lambda(\Phi^{\dagger} \Phi)^{2},
\label{L-model}
\ea
where $\Phi$ is a complex doublet field.
The 
chemical potential $\mu$ is provided by external conditions (to be specific, 
we take $\mu > 0$). For example, in the case of dense QCD with the kaon 
condensate, $\mu = m^{2}_s/2p_F$ where $m_s$ is the current mass of the 
strange quark and $p_F$ is the quark Fermi momentum \cite{BS}.
Note that the terms with the chemical potential
reduce the initial $SU(2)_{L}\times SU(2)_R$ symmetry   
to the $SU(2)_L \times U(1)_Y$ one. This follows from the fact that
the hypercharge generator $Y$ is $Y = 2I^{3}_{R}$ where $I^{3}_{R}$ 
is the
third component of the right handed isospin generator. Henceforth
we will omit the subscripts $L$ and $R$, allowing various
interpretations of the $SU(2)$ [for example, in the dynamics of
the kaon condensate, it is just the conventional
isospin symmetry $SU(2)_{I}$ and $\Phi^{T} = (K^{+},K^{0}$)].

The terms containing the chemical potential in Eq.~(\ref{L-model}) 
are
\ba
i\mu \Phi^{\dagger}\partial_{0}\Phi - i\mu\partial_{0}
\Phi^{\dagger}\Phi + \mu^{2}\Phi^{\dagger} \Phi. 
\label{mu-terms}
\ea
The last term in this expression makes the mass term in Lagrangian
density
(\ref{L-model}) to be $(\mu^{2} - m^2)\Phi^{\dagger} \Phi$. Therefore
for supercritical values of the chemical potential,
$\mu^2 > m^2$, there is an instability resulting in the spontaneous
breakdown of $SU(2) \times U(1)_Y$ down to 
$U(1)_{em}$ connected with the electrical charge 
$Q_{em} =  I^{3} + \frac{1}{2}Y$. One may expect that
this implies the existence of three
NG bosons. However, as was shown in Refs.~\cite{MS,STV}, 
there are only two NG bosons, which carry the quantum numbers of
$K^{+}$ and $K^{0}$ mesons. The third would-be NG boson, with
the quantum numbers of $K^{-}$, is massive in this model. This
happens despite the fact that the potential part
of Lagrangian (\ref{L-model}) has three flat directions in the broken
phase, as it should. The splitting between $K^{+}$ 
and $K^{-}$ occurs because of the seesaw mechanism in the kinetic 
part of the Lagrangian density (kinetic
seesaw mechanism) \cite{MS}. This mechanism is provided by the
first two terms in expression (\ref{mu-terms}) which, because of
the imaginary unit in front, mix the real and imaginary parts of the
field $\Phi$. Of course this effect is possible only because the $C$,
$CP$, and $CPT$ symmetries are explicitly broken in this system
at a nonzero $\mu$.

Another noticeable point is that while the dispersion relation for 
$K^{0}$ is conventional, with the energy $\omega \sim k$ as the momentum
$k$ goes to zero, the dispersion relation for 
$K^{+}$ is $\omega \sim k^2$ for small $k$ \cite{MS,STV}. This fact 
is in accordance 
with the Nielsen-Chadha counting rule, 
$N_{G/H} = n_{1} + 2n_{2}$ \cite {NC}. Here
$n_{1}$ is the number of NG bosons with the linear dispersion law,
$\omega \sim k$, $n_{2}$ is the number of NG bosons with the quadratic 
dispersion law $\omega \sim k^2$, and $N_{G/H}$ is the number of the 
generators in the coset space $G/H$ (here $G$ is the symmetry group of 
the action and $H$ is the symmetry group of the ground state).

Does the conventional Anderson-Higgs mechanism survive in the
gauged version of this model despite the absence of one out of 
three NG bosons? This question has motivated the present work.

\section{Gauged $\sigma$-model}

We will consider the  dynamics in the gauged version of model 
(\ref{L-model}), i.e., the model described by the Lagrangian
density
\ba
{\cal L}=-\frac{1}{4}F_{\mu\nu}^{(a)}F^{\mu\nu(a)}-\frac{1}{4}B_{\mu\nu}
B^{\mu\nu}+[(D_\mu-i\mu\delta_{\mu0})
\Phi]^\dagger(D^\mu-i\mu\delta^{\mu0})\Phi-m^2\Phi^\dagger\Phi-\lambda
(\Phi^\dagger\Phi)^2,
\lb{lagrangian}
\ea 
where the covariant derivative $D^\mu=\partial_\mu-igA_\mu- (ig^\prime/2)
B_\mu$, and 
\be
\Phi=\left(\begin{array}{c} 0 \\ \varphi_{0}
\end{array}\right)
+\frac{1}{\sqrt{2}}\left(\begin{array}{c}
\varphi_{1}+i\varphi_{2} \\
\tilde{\varphi}_{1}+i\tilde{\varphi}_{2}
\end{array}\right)
\ee
with $\varphi_0$ being the ground state expectation value.
The $SU(2)$ gauge fields are given by $A_\mu=A_\mu^a
\tau^a/2$, where $\tau^a$ are three Pauli matrices, and the
field strength $F_{\mu\nu}^{(a)}=\partial_\mu A_\nu^{(a)}-\partial_\nu 
A_\mu^{(a)}+g\epsilon^{abc}A_\mu^{(b)}A_\nu^{(c)}$. 
$B_\mu$ is the U(1)$_Y$
gauge field with the strength $B_{\mu\nu}=\partial_\mu B_\nu-\partial_\nu
B_\mu$. The hypercharge of the doublet $\Phi$ equals +1.
This model has the same structure as the electroweak theory without
fermions and with the chemical potential for hypercharge $Y$.

We will consider two different cases: the case with $g^{\prime} =0$, when
the hypercharge $Y$ is connected with the global $U(1)_{Y}$ symmetry,
and the case with a nonzero $g^{\prime}$, when the $U(1)_{Y}$ symmetry 
is gauged. The main results derived in this paper are the following.
For $m^2 > 0$, the spontaneous breakdown of the
$SU(2) \times U(1)_Y$ symmetry is caused solely by a
supercritical chemical potential $\mu^2 > m^2$. We show that
spontaneous breakdown of the $SU(2) \times U(1)_Y$ is
always accompanied by spontaneous breakdown of both the rotational
symmetry $SO(3)$ [down to $SO(2)$] and the electromagnetic $U(1)_{em}$ 
connected with the electrical charge. Therefore, in this case the
$SU(2) \times U(1)_Y \times SO(3)$ group is broken spontaneously 
down to $SO(2)$. This pattern of spontaneous symmetry breakdown
takes place for both $g^{\prime} =0$ and $g^{\prime} \neq 0$,
although the spectra of excitations in these two cases are different.
Also, the phase transition at the critical point $\mu^2 =m^2$ is
a second order one.

The realization of both the 
NG mechanism and the Anderson-Higgs mechanism
is conventional, despite the unconventional realization 
of the NG mechanism in the original ungauged model (\ref{L-model}).
For $g^{\prime} =0$, there are three NG
bosons with the dispersion
relation $\omega \sim k$,
as should be
in the conventional realization of the breakdown
$SU(2) \times U(1)_Y \times SO(3) \to SO(2)$ when 
$U(1)_Y \times SO(3)$ is a global symmetry. The other
excitations are massive (the Anderson-Higgs mechanism). 
For $g^{\prime} \neq 0$, there are two  
NG bosons with $\omega \sim k$, as should be
when only $SO(3)$ is a global symmetry (the third NG boson
is now ``eaten" by a photon like combination of fields
$A^{3}_{\mu}$ and $B_{\mu}$ 
that becomes massive). In accordance with the
Anderson-Higgs mechanism, the rest of excitations are massive. 

Since the residual $SO(2)$ symmetry is low, the spectrum
of excitations is very rich. In particular, the dependence of their
energies on the longitudinal momentum $k_3$, directed along the
$SO(2)$ symmetry axis, and on the transverse one, 
${\bf k}_\perp = (k_1,k_2)$, is quite different.
A noticeable point is
that there are two excitation branches, connected with two NG bosons,
that
behave as phonon like quasiparticles for small momenta
(i.e., their energy $\omega \sim k$) and
as roton like ones for large momenta $k_3$, i.e., there is a local minimum
in $\omega(k_3)$ for a value of $k_3$ of order $m$
(see Figs.~\ref{fig-all-0} and \ref{fig-all} below).
On the other hand, $\omega$ is a monotonically increasing function 
of the transverse momenta. The existence of the roton like
excitations is caused by the presence of gauge fields [there are
no such excitations in ungauged model (\ref{L-model})]. 
As is well known, excitations with the behavior of such a 
type are present in
superfluid systems \cite{superfluid}. This suggests that the present model
could be relevant for anisotropic superfluid systems.

In the case of $m^2 < 0$, the spontaneous breakdown of
the $SU(2) \times U(1)_Y$ symmetry takes place even without
chemical potential. Introducing the chemical potential
leads to dynamics similar to that in tumbling gauge
theories \cite{tumbling}. While in tumbling gauge
theories the initial symmetry is breaking down (``tumbling'')
in a few stages with 
increasing the running gauge coupling, in this model two different
stages of symmetry breaking are determined by the values of
chemical potential. When $0 < \mu^2 < \frac{g^2}{16\lambda}|m^2|$,
the $SU(2) \times U(1)_Y$ breaks down to $U(1)_{em}$, and the
rotational $SO(3)$ is exact. In this case, the conventional
Anderson-Higgs mechanism is realized with three gauge bosons being
massive and with no NG bosons.
The presence of $\mu$ leads to splitting of the masses of charged
$\pm 1$ gauge bosons. 

The second stage happens when 
$\mu^2$ becomes larger than $\frac{g^2}{16\lambda}|m^2|$. Then
one gets the same breaking sample as that described above
for $m^2 >0$, with $SU(2) \times U(1)_Y \times SO(3) \to SO(2)$.
The spectrum of excitations is also similar
to that case. 
At last, for all those values of the coupling constants $\lambda$ and $g$
for which the effective potential is bounded from below, the
phase transition at the critical 
point $\mu^2 = \frac{g^2}{16\lambda}|m^2|$ is a second order one.

\section{Model with global $U(1)_Y$ symmetry: $m^2>0$ case}

Before starting our analysis, we would like to make
the following general observation.
Let us consider a theory with a chemical potential $\mu$ 
connected with a conserved charge $Q$. Let us introduce
the quantity
\ba
R_{min} \equiv \mbox{min}(m^2/Q^2),
\lb{R}
\ea
where on the right hand side we consider the minimum value 
amongst the ratios
$m^2/Q^2$ for all {\it bosonic} particles with $Q \neq 0$
in this same theory but
{\it without} the chemical potential. Then if $\mu^2 > R_{min}$,
the theory exists only if the spontaneous breakdown 
of the $U(1)_{Q}$ symmetry takes place there. 
Indeed, if the $U(1)_{Q}$ were exact in such a theory,
the partition function, $Z = \rm{Tr}[exp(\mu \hat{Q} - H)/T]$, 
would diverge
\footnote{Similarly as it happens in nonrelativistic bose gas with a
positive chemical potential connected with the number
of particles $N$ \cite{LL}.}. 
Examples of the restriction $\mu^2 < R_{min}$ in relativistic
theories were considered
in Refs.~\cite{Kapusta,HW}. 

In fact, the value $\mu^2 = R_{min}$
is a critical point separating different phases in the theory.
It is important that since in the phase
with $\mu^2 > R_{min}$ the charge $Q$ is not a good quantum
number, $\mu$ ceases to play the role of a chemical
potential determining the density of this charge.
This point was emphasized in Ref. \cite{Kapusta}.
There are a few options in this case. If there remains an exact symmetry
connected with a charge $Q^{\prime} = aQ + X$, where 
$a$ is a constant and $X$ represents some other generators,
the chemical
potential will determine the density of the charge $Q^{\prime}$
(a dynamical transmutation of the chemical potential).
Otherwise, it becomes just a
parameter determining the spectrum of excitations and other
thermodynamic
properties of the system (the situation is similar to that taking place
in models when a mass square $m^2$ becomes negative). We will
encounter both these options in model (\ref{lagrangian}).

We begin by considering the case with $g^{\prime} = 0$ 
and $m^2 > 0$. When $\mu^{2} < m^2$, the
$SU(2) \times U(1)_Y \times SO(3)$ symmetry is exact. Of course
in this case a confinement dynamics for three $SU(2)$
vector bosons takes place and it is not under our control. However,
taking $\mu^2 \sim m^2$ and choosing $m$ to be much larger than the
confinement scale $\Lambda_{SU(2)}$, we get controllable
dynamics at large momenta $k$ of order $m$. It includes three
massless vector bosons $A^{a}_{\mu}$ and two doublets,
$(K^{+},K^{0})$ and $(K^{-}, \bar{K}^{0})$. The spectrum of
the doublets is qualitatively the same as that in model
(\ref{L-model}): the chemical potential leads to splitting
the masses (energy gaps) of these doublets and, in tree
approximation, their masses are $m - \mu$ and $m + \mu$,
respectively \cite{MS,STV}. In order to make the tree
approximation to be reliable, one should take $\lambda$
to be small but much larger than the value
of the running coupling $g^4(m)$ related to the scale $m$
[smallness of $g^2(m)$ is guaranteed by the condition
$m \gg \Lambda_{SU(2)}$ assumed above]. The condition
$g^4(m)\ll \lambda \ll 1$ implies that 
the contributions both of vector boson
and scalar loops are small, i.e., there is no Coleman-Weinberg (CW)
mechanism (recall that one should have
$\lambda \sim g^4$ for the CW mechanism) \cite{CW}.         

Let us now consider the case with $\mu^2 > m^2 >0$ in detail. 
Since $m^2$ is equal to $R_{min}$ (\ref{R}), there should be 
spontaneous $U(1)_{Y}$ symmetry breaking in this case.   
For $g^{\prime}= 0$,
the equations of motion derived from Lagrangian density  
Eq.(\ref{lagrangian}) read:
\ba
&&-(D_\mu-i\mu\delta_{\mu0})(D^\mu-i\mu\delta^{\mu0})\Phi-m^2\Phi-2\lambda
(\Phi^\dagger\Phi)\Phi=0,
\label{equation1}\\
&&\partial^\mu F_{\mu\nu}^{(a)}+g\epsilon^{abc}A^{\mu(b)} F_{\mu\nu}^{(c)}
+ig\left[\Phi^\dagger\frac{\tau^a}{2}\partial_\nu\Phi-\partial_\nu
\Phi^\dagger\frac{\tau^a}{2}\Phi\right]+\frac{g^2}{2}A_{\nu}^{(a)}
\Phi^\dagger\Phi
+2g\mu\delta_{\nu0}\Phi^\dagger\frac{\tau^a}{2}\Phi=0
\label{equation2}
\ea
(since now the field $B_{\mu}$ is free and decouples, we ignore it).
Henceforth we will use the unitary gauge with 
$\Phi^T=(0,\varphi_0+\tilde\varphi_1/\sqrt{2})$.
It is important that the existence of this gauge is based solely on
the presence of $SU(2)$ gauge symmetry, independently of
whether the number of
NG bosons in ungauged model (\ref{L-model}) is conventional or not. 
We will be first
looking for a homogeneous ground state solution (with $\varphi_0$ 
being constant) that does not break the
rotational invariance, i.e., with $A_i^{(3,\pm)}=0$ where
$A_\mu^{(\mp)}=\frac{1}{\sqrt{2}}(A_\mu^{(1)}\pm iA_\mu^{(2)})$.
In this case the equations of motion become
\ba
\left(i\partial_0A_0^{(+)}+2\mu A_0^{(+)}\right)
\varphi_0=0,&&\\
\left[(\mu-\frac{g}{2}A_0^{(3)})^2-m^2-
2\lambda\varphi_0^2-\frac{ig}{2}
\partial_0A_0^{(3)}+\frac{g^2}{2}A_0^{(+)}A_0^{(-)}\right]
\varphi_0=0,&&\\
g\left(\frac{g}{2}A_0^{(3)}-\mu\right)\varphi_0^2
=0,&&\\
\frac{g^2\varphi_0^2}{2}A_0^{(\pm)}=0.
\ea
Besides the symmetric solution with $\varphi_0=0$,
this system of equations allows the following solution
\ba
\varphi_0^2=-\frac{m^2}{2\lambda},\quad
A_0^{(3)}=\frac{2\mu}{g},\quad A_0^{(\pm)}=0.
\label{vacuum2}
\ea
We recall that in the unitary gauge
all auxiliary, gauge dependent, degrees of
freedom are removed. Therefore in this gauge
the ground state expectation values of vector
fields are well defined physical quantities.

Solution (\ref{vacuum2}), describing spontaneous $U(1)_{Y}$ symmetry
breaking, exists only for negative $m^2$. On the other hand, the symmetric
solution with $\varphi_0=0$
cannot be stable in the case of
$\mu^2 > R_{min}= m^2 >0$ we are now interested in. This forces us to
look for a ground state solution that breaks the rotational 
invariance \footnote{We will get a better insight in the reason why
spontaneous rotational invariance breaking is inevitable for
$\mu^2 > m^2 >0$ from considering the dynamics with
$m^2 < 0$ below.}. Let us now consider the effective potential $V$.
It is obtained from Lagrangian density 
Eq.(\ref{lagrangian}), $V=-{\cal L}$, by setting all field 
derivatives to zero. Then we get:
\ba
V=V_1+V_2,
\lb{potential}
\ea
with
\ba
V_1&=&-\frac{g^2}{2}\left[\left(A_0^{(a)}A_0^{(a)}\right)\left(A_i^{(b)}
A_i^{(b)}\right)-
\left(A_0^{(a)}A_i^{(a)}\right)\left(A_0^{(b)}A_i^{(b)}\right)\right]
\nonumber\\
&+&\frac{g^2}{4}\left(A_i^{(a)}A_i^{(a)}
\right)\left(A_j^{(b)}A_j^{(b)}\right)-\frac{g^2}{4}\left(A_i^{(a)}A_j^{(a)}
\right)\left(A_i^{(b)}A_j^{(b)}\right),\\
V_2&=&(m^2-\mu^2)\Phi^\dagger\Phi+\lambda(\Phi^\dagger\Phi)^2
-2g\mu\Phi^\dagger A_0^{(a)}\frac{\tau^a}{2}\Phi-\frac{g^2}{4}
A_\mu^{(a)}A^{\mu(a)}\Phi^\dagger\Phi.
\ea
We use the ansatz 
\be
A_3^{(+)}=(A_3^{(-)})^*=C\neq0,\quad A_0^{(3)}=D\neq0,
\quad A_{1,2}^{(\pm)}
=A_{0}^{(\pm)}=A_{1,2}^{(3)}=A_{3}^{(3)}=0,\quad \Phi^T=(0,\varphi_{0}) 
\label{ansatz} 
\ee 
that breaks 
spontaneously both rotational symmetry [down to $SO(2)$] and
$SU(2)\times U(1)_Y$ [completely]. Substituting
this ansatz into potential (\ref{potential}), we
arrive at the expression 
\be
V=-g^2D^2|C|^2-\left(\mu-\frac{gD}{2}\right)^2
\varphi_0^2+\frac{g^2}{2}|C|^2\varphi_0^2+m^2\varphi_0^2+\lambda\varphi_0^4.
\label{potential-with-anzats}
\ee
It leads to the following equations of motion:
\ba
\left(D^2-\frac{\varphi_0^2}{2}\right)C=0,&&\\
\left(2|C|^2+\frac{\varphi_0^2}{2}\right)D=\frac{\varphi_0^2}{g}
\mu,&&\\
\left[\left(\mu-\frac{gD}{2}\right)^{2} -m^2-2\lambda
\varphi_0^2-\frac{g^2}{2}|C|^2 \right]\varphi_0=0.
\label{anzats-eq}
\ea
One can always take both $g$ and the ground state expectation
value $\varphi_0$ to be positive (recall that we also
take $\mu>0$). Then from the first two equations we obtain
\be
 D=\frac{\varphi_0}{\sqrt{2}}>0,\quad 2|C|^2+\frac{\varphi_0^2}{2}=
\frac{\sqrt{2}\mu\varphi_0}{g},
\label{solution-for-A03}
\ee
while the third equation reduces to 
\be
\left(\frac{g^2}{4}-2\lambda\right)\varphi_0^2-\frac{3g\mu}{2\sqrt{2}}
\varphi_0+\mu^2-m^2=0.
\label{equation}
\ee
Hence for $\varphi_0$ we get the following solution
\be
\varphi_0=\frac{1}{\sqrt{2}(8\lambda-g^2)}\left[\sqrt{(g^2+64\lambda)
\mu^2-8(8\lambda-g^2)m^2}-3g\mu\right].
\label{solution-varphi}
\ee
It is not difficult to show that for $\mu^2 > m^2 > 0$ both expression 
(\ref{solution-varphi}) for $\varphi_0$ and expression 
(\ref{solution-for-A03}) for $|C|^2$ are positive and, 
for $g^2\leq 8\lambda$, this solution corresponds to the minimum
of the potential. The phase transition at the critical value 
$\mu = m$ is a second order one. 

The situation in the region $g^2> 8\lambda$ is somewhat more
complicated. First of all, in that region the potential 
(\ref{potential-with-anzats}) becomes unbounded from below 
[one can see this after substituting the expression for 
$A_0^{(3)}=D$ from Eq.~(\ref{solution-for-A03}) into the potential]. 
Still, even in that case there is
a local minimum corresponding to solution (\ref{solution-varphi}).
The phase transition is again a second order one. Henceforth
we will consider only the case with $g^2\leq 8\lambda$ when the
potential is bounded from below. 
Notice that for small $g^2 \equiv g^2(m)$ the inequality
$g^2\leq 8\lambda$ is consistent with the condition
$g^4 \ll \lambda$ necessary for the suppression of the
contribution of vector boson loops, as was discussed above.

In order to study the spectrum of excitations, we take 
for convenience $C$ to be 
positive and make the expansion in Lagrangian
density (\ref{lagrangian}) about the ground state solution 
in Eq.~(\ref{ansatz}). Introducing small fluctuations 
$a^{(a)}_{\mu}$ 
[i.e., $A^{(a)}_{\mu} = \langle A^{(a)}_{\mu} \rangle + a^{(a)}_{\mu}$] 
and $\tilde\varphi_1$
[i.e., $\Phi^T=(0,\varphi_0+\tilde\varphi_1/\sqrt{2})]$ 
and keeping only quadratic fluctuation terms, we get:
\be
{\cal L}={\cal L}_{(i0)}+{\cal L}_{(ij)}+ 
{\cal L}_{(\varphi)},
\label{form}
\ee
where
\ba
{\cal L}_{(i0)}&=&\frac{1}{2}f_{i0}^{(a)}f_{i0}^{(a)}+gD\left(f_{i0}^{(1)}
a_i^{(2)}-f_{i0}^{(2)}a_i^{(1)}\right)+\sqrt{2}gC\left(f_{30}^{(3)}
a_0^{(2)}-f_{30}^{(2)}a_0^{(3)}\right)+g^2C^2\left(a_0^{(2)}a_0^{(2)}+
a_0^{(3)}a_0^{(3)}\right)\nonumber\\
&+&\sqrt{2}g^2CD\left(2a_0^{(3)}a_3^{(1)}-
a_0^{(1)}a_3^{(3)}\right)+\frac{g^2D^2}{2}\left(a_i^{(1)}a_i^{(1)}+
a_i^{(2)}a_i^{(2)}\right),\\
{\cal L}_{(ij)}&=&-\frac{1}{4}f_{ij}^{(a)}f_{ij}^{(a)}-\sqrt{2}gC\left(
f_{i3}^{(2)}a_i^{(3)}-f_{i3}^{(3)}a_i^{(2)}\right)-g^2C^2\left(
a_1^{(2)}a_1^{(2)}+a_1^{(3)}a_1^{(3)}+a_2^{(2)}a_2^{(2)}+a_2^{(3)}
a_2^{(3)}\right),\\
{\cal L}_{(\varphi)}&=&\frac{1}{2}\partial_\mu\tilde\varphi_1
\partial^\mu\tilde\varphi_1
-\frac{1}{2}\left[m^2-\left(\mu-\frac{gD}{2}\right)^2+\frac{g^2C^2}{2}
+6\lambda\varphi_0^2\right]\tilde\varphi_1^2
+\sqrt{2}g\left(\frac{gD}{2}-\mu\right)\varphi_0\tilde\varphi_1 a_0^{(3)}
\nonumber\\
&-&g^2C\varphi_0\tilde\varphi_1 a_3^{(1)}
+\frac{g^2}{4}\varphi_0^2a_\mu^{(a)}a^{\mu (a)}
\ea
with $f_{\mu\nu}^{(a)}=\partial_\mu a^{(a)}_\nu-
\partial_\nu a^{(a)}_\mu$.

Since in the subcritical phase, with $\mu^2 < m^2$, there 
are 10 physical states (6 states connected with three 
massless vector bosons and 4 states connected with the
doublet $\Phi$), one should expect that there should
be 10 physical states (modes) also in the supercritical
phase described by the quadratic form (\ref{form}).
The analysis of this quadratic form was done by using 
{\it MATHEMATICA}. 
It leads to the following spectrum of
excitations. Out of the total 10 modes there exist 3
massless (gapless) NG modes, as should be in the conventional
realization of the spontaneous breakdown of
$SU(2) \times U(1)_Y \times SO(3) \to SO(2)$, when
$U(1)_Y \times SO(3)$ is a global symmetry. The gaps (``masses")
$\Delta$ of the excitations are defined as the values of their
energies at zero momentum. They are:
\be
\begin{array}{ll}
\Delta^2 = 0, & [\times 3] ,\\
\Delta^2 = 2 \mu \phi, & [\times 2] ,\\
\Delta^2 = 2 \mu \phi + 3\phi^2 , & [\times 2] ,\\
\Delta^2 = 4 \mu^2,  & [\times 1] ,\\
\Delta^2 = \delta_{-}^2,& [\times 1] ,\\
\Delta^2 = \delta_{+}^2, & [\times 1] 
\end{array}
\label{gap}
\ee
with the degeneracy factors specified in square brackets. Here
we introduced the following notations:
\ba
\phi^2 = g^2\varphi_0^2/2\quad \mbox{and} \quad
\delta_{\pm}^2 = F_1\pm\sqrt{F_1^2 - F_2}
\ea
with
\ba
F_1 &=& 3\mu^2-m^2 -\frac{7}{2}\mu \phi + 3\phi^2,\\
F_2  &=& 8(3\mu^2-m^2)\phi^2-30\mu \phi^3+9\phi^4.
\ea

The dispersion relations for the NG bosons in the infrared
region are:  
\ba
\omega^2 &\simeq& \frac{2\mu-\phi}{2\mu+3\phi}{\bf k}^2+O(k_i^4),\\
\omega^2 &\simeq& \frac{2\mu-\phi}{2\mu+3\phi}\left(
\frac{\phi}{2\mu}{\bf k}_\perp^2+k_3^2\right)+O(k_i^4),\\ 
\omega^2 &\simeq& \frac{ \left(2\mu-\phi\right) \left[
4(\mu^2-m^2)-3\mu\phi\right]}
{\mu \left[ 8(3\mu^2-m^2) -30 \mu\phi+9\phi^2\right]}{\bf k}^2
+O(k_i^4),
\ea
where $\omega \equiv k_0$. The infrared dispersion relations for the other
seven excitations read ($|k_i|\ll\phi$)
\ba
\omega^2 &\simeq& 2\phi \mu +{\bf k}_\perp^2+\frac{7\phi-8\mu}{3\phi}k_3^2
+O(k_i^4),\\
\omega^2 &\simeq& 2\phi \mu
+\frac{4\mu^2(3\mu^2-m^2)-22\mu^3\phi+2(7\mu^2+2m^2)\phi^2}
{4\mu^2(3\mu^2-m^2)-2\mu\phi(21\mu^2-4m^2)+42\mu^2\phi^2-9\mu\phi^3}
{\bf k}_\perp^2
+\frac{7\phi-8\mu}{3\phi}k_3^2+O(k_i^4),\\
\omega^2 &\simeq& 2\phi \mu +3\phi^2
+\frac{2\mu+7\phi}{2\mu+3\phi}{\bf k}_\perp^2
+ \frac{16\mu^2+22\mu\phi+9\phi^2}{3 \phi \left(2\mu+3\phi\right)}k_3^2
+O(k_i^4),\\
\omega^2 &\simeq& 2\phi \mu +3\phi^2
+4\frac{2\mu^3-\mu^2\phi-6\mu\phi^2-3\phi^3}
{8\mu^3+8\mu^2\phi-12\mu\phi^2-9\phi^3}{\bf k}_\perp^2
+ \frac{16\mu^2+22\phi\mu+9\phi^2}{3 \phi \left(2\mu+3\phi\right)}k_3^2
+O(k_i^4),\\
\omega^2 &\simeq& 4\mu^2 +\frac{16\mu^3-8\mu\phi^2-\phi^3}
{2\mu(4\mu^2-2\mu\phi-3\phi^2)}{\bf k}_\perp^2 \nonumber \\
&+&\frac{16\mu^3(\mu^2-m^2)
-4\mu^2(19\mu^2+3m^2)\phi
-2\mu(9\mu^2-4m^2)\phi^2
+2(31\mu^2+2m^2)\phi^3  
-15\mu\phi^4}{\mu\left[
8\mu^2(\mu^2-m^2) -28\mu^3\phi +8m^2\phi^2 +30\mu\phi^3-9\phi^4
\right]} k_3^2 + O(k_i^4),\\
\omega^2 &\simeq& \delta_{-}^2 + v_{\perp}^2 {\bf k}_\perp^2
+ v_{3}^2 k_3^2 + O(k_i^4),\\
\omega^2 &\simeq& \delta_{+}^2 + w_{\perp}^2 {\bf k}_\perp^2
+ w_{3}^2 k_3^2 + O(k_i^4),
\ea
where $v_{\perp}$, $v_{3}$, $w_{\perp}$ and $w_{3}$ are rather
complicated functions of the parameters $\mu$, $m$ and $\phi$.

While the analytical dispersion relations in the infrared
region are
quite useful, we performed also numerical calculations 
to extract the corresponding dispersion relations outside
the infrared region. The results are as follows. 

\begin{figure}
\mbox{
\includegraphics[bbllx=88,bblly=4,bburx=588,bbury=313,width=8cm]
{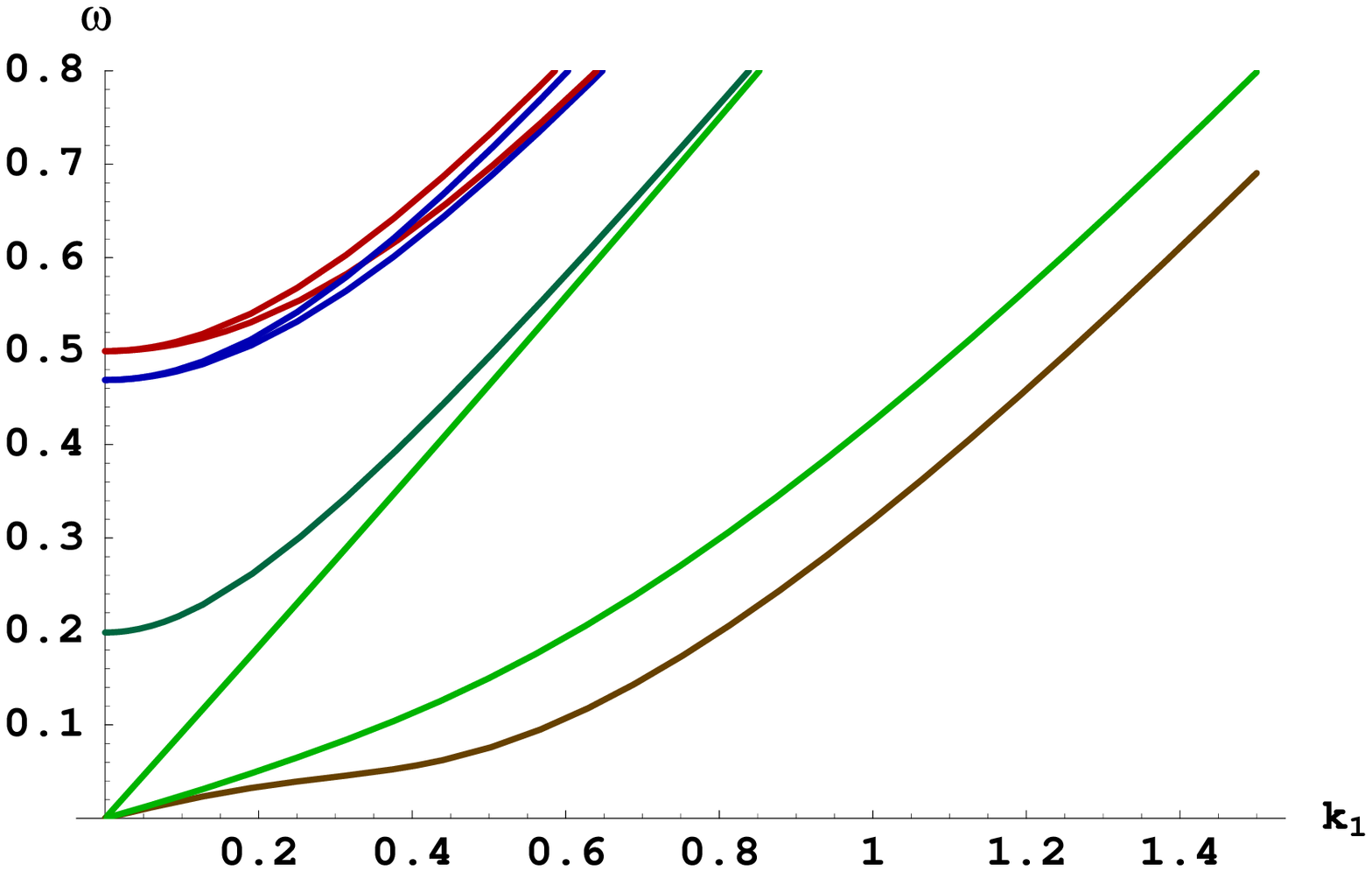}
\includegraphics[bbllx=88,bblly=4,bburx=588,bbury=313,width=8cm]
{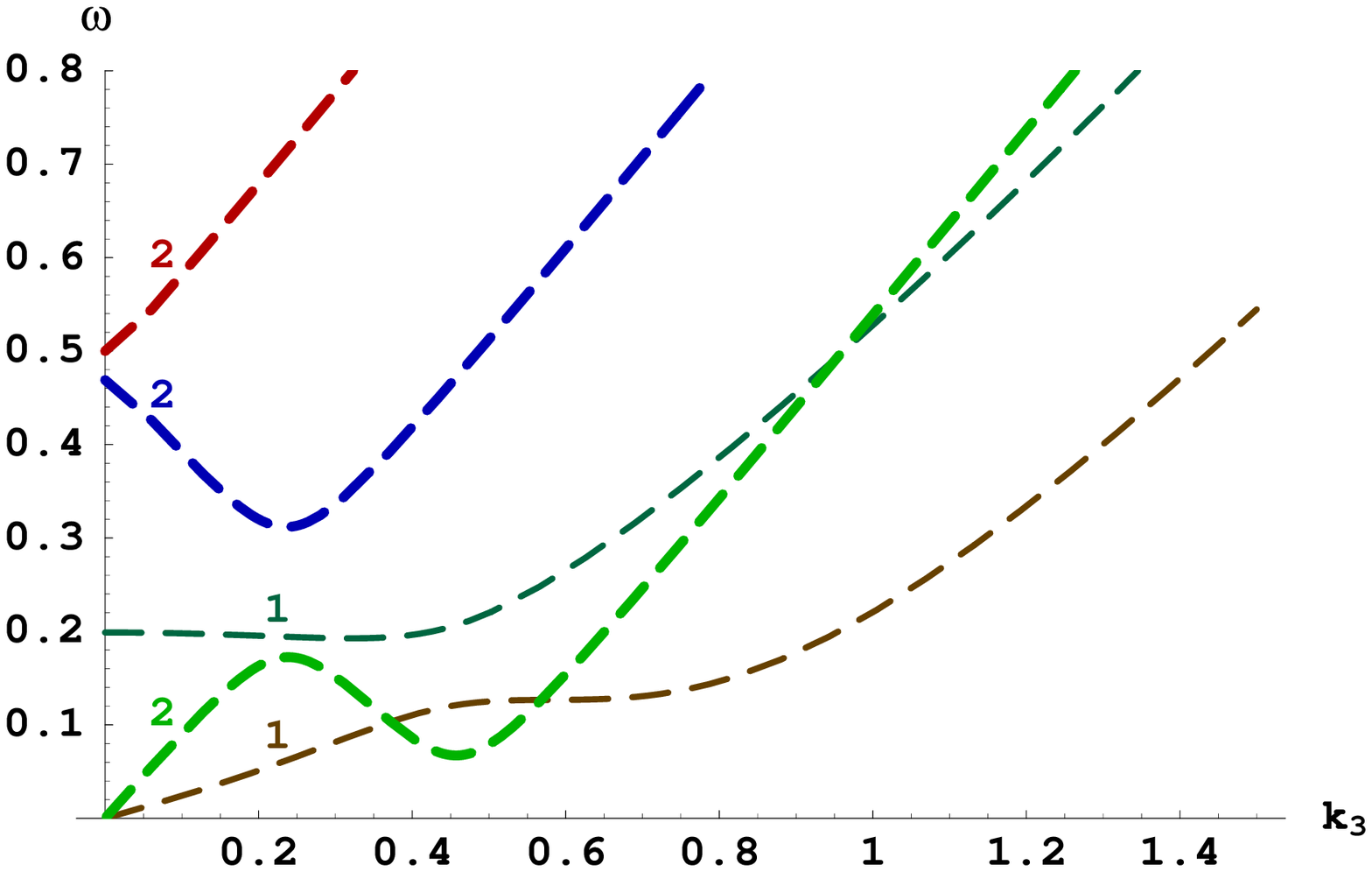}}

\caption{The energy $\omega$ of the 8 light quasiparticle modes as a
function of $k_1$ (solid lines, left panel) and $k_3$ (dashed lines,
right panel). The dispersion relations of two heavy modes are outside
the plot range. The energy and momenta are measured in units of $m$.
The parameters are $\mu/m=1.1$ and $\phi/m=0.1$}

\label{fig-all-0}
\end{figure}

In the near-critical region, $\mu\to m+0$, the ground state expectation 
$\phi$ becomes small. In this case, one gets 8 light modes,
see Eq. (\ref{gap}). 
The 
results for their dispersion relations are shown 
in Fig.~\ref{fig-all-0}
[the two heavy modes with the gaps of order $2\mu$
are not shown there].
The solid and dashed lines represent the energies
of the quasiparticle modes as functions of the transverse
momentum ${\bf k}_\perp = (k_1,0)$ (with $k_3 =0$) 
and the longitudinal momentum $k_3$ (with ${\bf k}_\perp =0$),
respectively. Bold and thin lines correspond 
to double degenerate and nondegenerate modes, respectively. 

There are the following characteristic features of the spectrum.
a) The spectrum with ${\bf k}_\perp =0$ (the right panel in 
Fig.~\ref{fig-all-0}) is much more degenerate 
than that with $k_3 =0$ (the left panel). This
point reflects the fact that the axis of the residual $SO(2)$
symmetry is directed along $k_3$. Therefore the states with
${\bf k}_\perp =0$ and $k_3 \neq 0$ are more symmetric than
those with ${\bf k}_\perp \neq 0$. b) The right panel in 
Fig.~\ref{fig-all-0} contains two 
branches with local minima at
$k_3 \sim m$, i.e., roton like excitations. Because there are no such
excitations in ungauged model (\ref{L-model}) \cite{MS,STV}, 
they occur because of the presence of gauge fields.
Since roton like excitations occur in superfluid systems, 
the present model could be relevant for them. c) The NG 
and Anderson-Higgs mechanisms are conventional in this system. 
In particular, the dispersion relations for three NG bosons
have the form $\omega \sim k$ for low momenta.

When the value of the chemical potential increases, the values of 
masses of all massive quasipatricles become of the same order.
Otherwise, the characteristic features of the dispersion
relations remain the same.

\section{Model with global $U(1)_Y$ symmetry: $m^2<0$ case}

Let us now turn to the case with negative $m^2$. In this case
there is the ground state solution (\ref{vacuum2}) describing
spontaneous breakdown of $SU(2)\times U(1)_{Y}$ down
to $U(1)_{em}$ and preserving the rotational invariance.
In order to describe the spectrum of excitations, 
we make the expansion in Lagrangian
density (\ref{lagrangian}) about this solution.
Introducing as before small fluctuations $a^{(a)}_{\mu}$
[i.e., $A^{(a)}_{\mu} = \langle A^{(a)}_{\mu} \rangle +a^{(a)}_{\mu}$] 
and $\tilde\varphi_1$
[i.e., $\Phi^T=(0,\varphi_0+\tilde\varphi_1/\sqrt{2})]$ 
and keeping only quadratic fluctuation terms, we obtain:
\ba
{\cal L} &\simeq &
\frac{1}{2}f_{0i}^{(a)}f_{0i}^{(a)}- \frac{1}{4}f_{ij}^{(a)}
f_{ij}^{(a)}+2\mu\left(f_{0i}^{(2)}a_i^{(1)}-f_{0i}^{(1)}a_i^{(2)}\right)
+2\mu^2\left(a_i^{(1)}a_i^{(1)}+a_i^{(2)}a_i^{(2)}\right)
+\frac{g^2\varphi_0^2}{4}a_\mu^{(a)}a^{\mu(a)}\nonumber \\
&+&\frac{1}{2}\partial_\mu\tilde{\varphi}_{1}
\partial^\mu\tilde{\varphi}_{1}
-\frac{1}{2}\left(m^2+6\lambda\varphi_{0}^2\right)\tilde{\varphi}_{1}^2.
\label{L}
\ea
The analysis of the spectrum of eigenvalues of this quadratic
form is straightforward. The dispersion relations for
charged vector bosons are:
\ba
\omega^{2}_{-}  &=& (\sqrt{{\bf k}^2 +\phi^2} + 2\mu)^2,\\
\omega^{2}_{+} &=& (\sqrt{{\bf k}^2 +\phi^2}- 2\mu)^2 ,
\label{charged}
\ea
where $\omega_{+}$ and $\omega_{-}$ are the energies of vector
bosons with $Q_{em} = +1$ and $Q_{em} = -1$, respectively. The
dispersion relations for the neutral vector boson 
and neutral scalars are
$\mu$ independent:
\ba
\omega^{2}_{0} & =& {\bf k}^2 +\phi^2,\\
\omega^{2}_\varphi& =& {\bf k}^2 +4\lambda\varphi_0^2.
\ea
Therefore, the chemical potential leads to splitting the
masses of two charged vector bosons. In fact, it is easy
to check that the terms with the chemical potential in
Lagrangian density (\ref{L}) look exactly as if the
chemical potential $\bar{\mu} = 2\mu$ for the electric
charge $Q_{em}$ was introduced. In other words, 
as a result of spontaneous $U(1)_Y$ symmetry breaking,
the dynamical transmutation of the chemical potential
occurs: the chemical potential for hypercharge transforms
into 
the chemical potential for electrical charge. Since the
hypercharge of vector bosons equals zero and $\tilde\varphi_1$
scalar is neutral, this transmutation looks quite dramatic:
instead of a nonzero density for scalars, a 
nonzero density for charged vector boson is generated.
(The factor
$2$ in $\bar{\mu} = 2\mu$ is of course connected with
the factor $1/2$ in $Q_{em} = I^3 + \frac{1}{2}Y$.)

In this phase, the parameter 
$R_{min}$ (\ref{R}) equals $\phi^2 =\frac{g^2}{4\lambda}|m^2|$,
i.e., it coincides with the square of the mass of vector
bosons in the theory without chemical potential.
Therefore, as the chemical potential $\bar{\mu}^2$ becomes larger
than $\phi^2 =R_{min}$, a new phase transition should happen.
And since for $\bar{\mu}^2 = \phi^2$ vector bosons
with charge $+1$ become gapless [see Eq. (\ref{charged})],
one should expect that
this phase transition is 
triggered by generating a condensate of
charged vector bosons. 

And such a condensate arises indeed. It is not
difficult to check that when 
$\bar{\mu}^2 > \bar{\mu}^{2}_{cr} \equiv  \frac{g^2}{4\lambda}|m^2|$ ,
the ground state solution with ansatz (\ref{ansatz})
occurs. The parameters $C$, $D$, and $\varphi_0$ are
determined from Eqs.~(\ref{solution-for-A03}) and
(\ref{solution-varphi}), respectively. For 
$\bar{\mu}^2 >\bar{\mu}^{2}_{cr}$,
both expression
(\ref{solution-varphi}) for $\varphi_0$ and
expression (\ref{solution-for-A03}) for
$|C|^2$ are positive and,
for $g^2\leq 8\lambda$, this solution corresponds to the 
global minimum
of the potential. The phase transition at the critical value
$\bar{\mu}^2 = \bar{\mu}^{2}_{cr}$
is a second order one \footnote{As was shown
above, for 
$g^2 > 8\lambda$,
the potential (\ref{potential-with-anzats}) is unbounded from
below, and we will not consider this case.}. 
The spectrum of excitations in the supercritical phase with
$\bar{\mu}^2 > \bar{\mu}^{2}_{cr}$
is similar to the spectrum in the
case of positive $m^2$ and $\mu^2 > m^2$ shown in Fig.~\ref{fig-all-0}.

Therefore, for $m^2 <0$ the breakdown of the initial symmetry 
is realized
in two steps, similarly as it takes place in tumbling gauge
theories \cite{tumbling}. 
Now we can understand more clearly 
why in the case of positive $m^2$ considered above
the breakdown of the initial symmetry is realized in
one stage. The point is that in that case 
vector bosons in the theory without chemical potential are
massless. Therefore, while $R_{min} =m^2 >0$ for the
chemical potential for hypercharge, 
$R_{min} =0$ for the chemical
potential connected with electrical charge $Q_{em}$ 
there. 
This in turn implies that in that case there is no way for 
increasing 
$R_{min}$ through the process of
the transmutation of the chemical potential as it happens
in the case of negative $m^2$. Therefore for $m^2 >0$
the phase in which both the
$U(1)_{em}$ symmmetry and the rotational symmetry
are broken occurs at once as $\mu^2$ becomes larger
than $m^2$. 





\section{Model with gauged $U(1)_Y$ symmetry}

Let us now briefly describe the case with $g^{\prime} \neq 0$.
In this case the $U(1)_{Y}$ symmetry is local and one should
introduce a source term $B_{0}J_{0}$ in Lagrangian density 
(\ref{lagrangian}) in order to make the system neutral
with respect to hypercharge $Y$. This is necessary since
otherwise in the system with a nonzero chemical potential
$\mu$ thermodynamic equilibrium could not be established.
The value of the background hypercharge density $J_{0}$
[representing very heavy particles] is determined from
the condition $\langle{B_0}\rangle =0$ \cite{Kapusta}.  

After that, the analysis follows closely to that of
the case with $g^{\prime} =0$.
Because of the additional vector boson $B_{\mu}$,
there are now 12 quasiparticles in the spectrum. 
The sample of spontaneous 
$SU(2) \times U(1)_Y \times SO(3)$ symmetry breaking
is the same as for $g^{\prime} =0$ both for $m^2 \geq 0$
and $m^2 < 0$, with a tumbling like scenario for the latter.
However, for supercritical values of
the chemical potential, there are now 
only two gapless NG modes [the third one is ``eaten" 
by a photon
like combination of fields $A^{3}_{\mu}$ and $B_{\mu}$ that becomes
massive].
Their dispersion relations in infrared read
\ba
\omega^2 &\simeq& \frac{2\mu-\phi}{2\mu+3\phi}{\bf k}^2+O(k_i^4),\\
\omega^2 &\simeq& \frac{2\mu-\phi}{2\mu+3\phi}\left(
\frac{\phi}{2\mu}{\bf k}_\perp^2+k_3^2\right)+O(k_i^4).
\ea
The rest 10 quasiparticles are gapped. The mass (gap) of the
two new states is:  
\ba
\Delta^2 = \mu \phi + \frac{\phi_b^2 }{2}
-\sqrt{\left(\mu \phi -\frac{\phi_b^2}{2}\right)^2+\phi^2 \phi_b^2},
& [\times 2],
\label{gap1}
\ea
where $\phi_b^2 =(g^{\prime})^2\varphi_0^2/2$
with $\varphi_0^2$ given in Eq. (\ref{solution-varphi}).
This gap goes to zero together with $g^{\prime}$, i.e., these two
degrees of freedom correspond to two transverse states
of massless vector boson $B_{\mu}$ in this limit. 

\begin{figure}


\mbox{
\includegraphics[bbllx=88,bblly=4,bburx=588,bbury=313,width=8cm]
{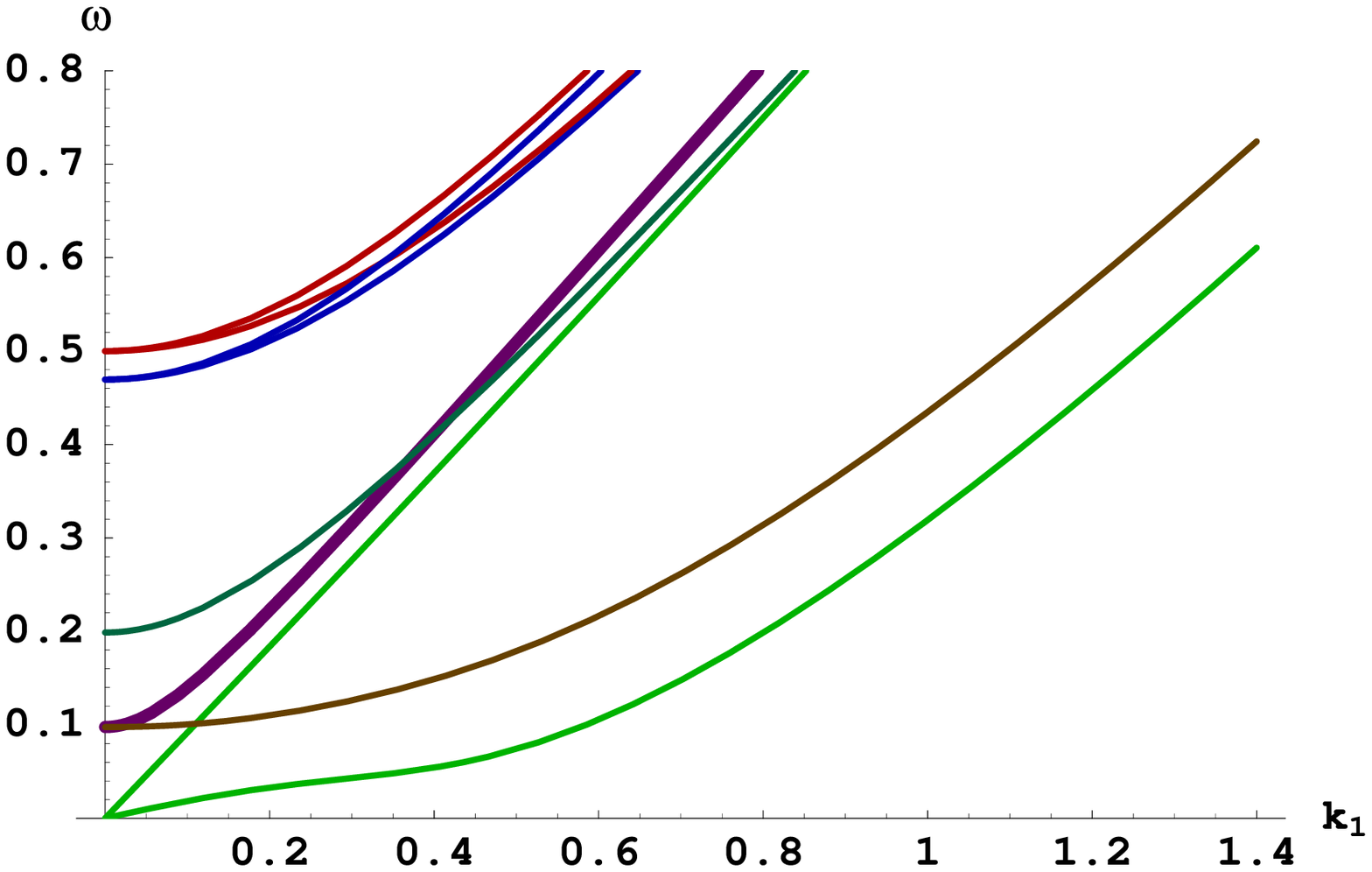}
\includegraphics[bbllx=88,bblly=4,bburx=588,bbury=313,width=8cm]
{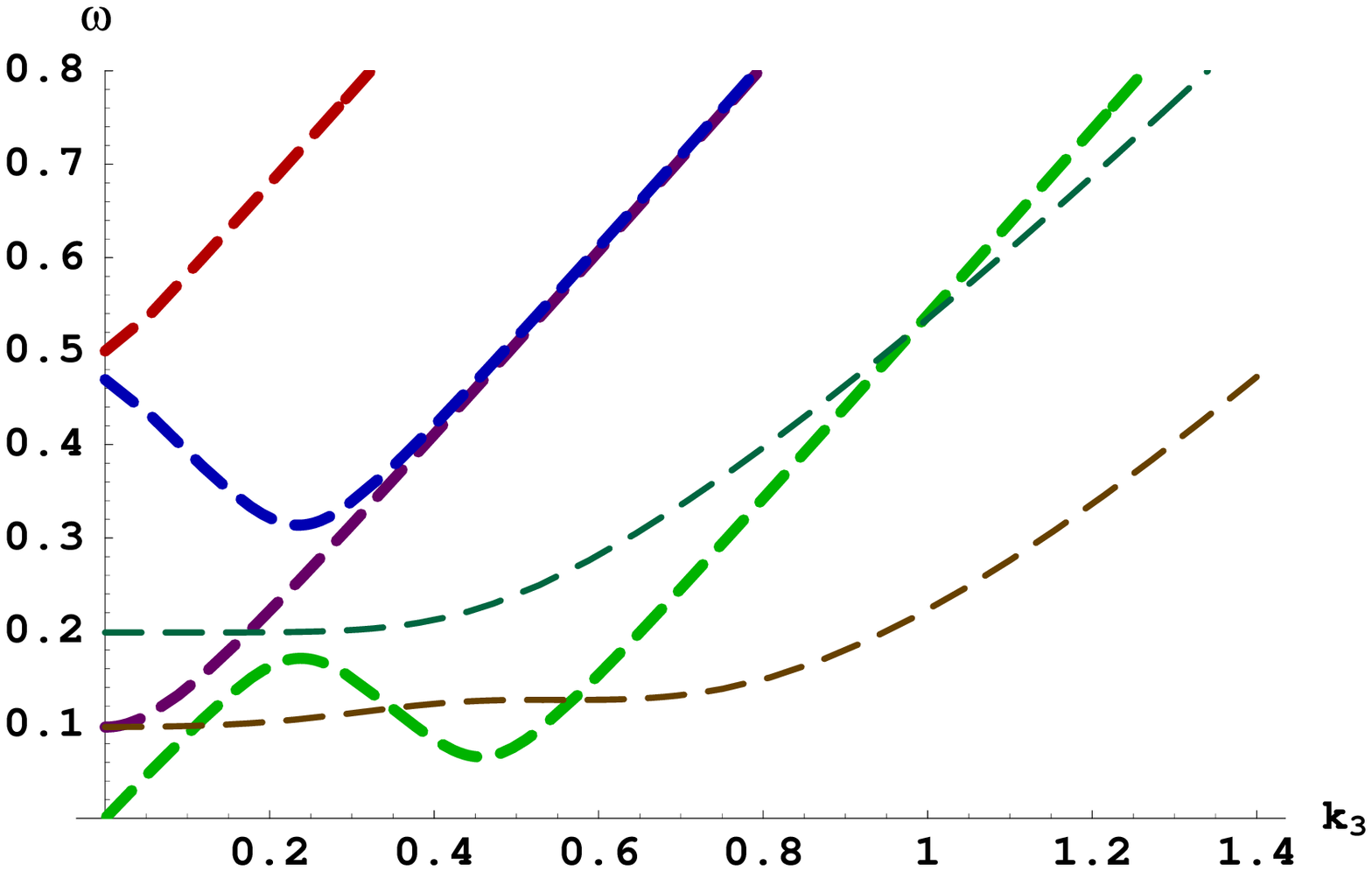}}

\caption{The energy $\omega$ of the 10 light quasiparticle modes as a 
function of $k_1$ (solid lines, left column) and $k_3$ (dashed lines,
right column). The dispersion relations of two heavy modes are outside
the plot ranges. The energy and momenta are measured in units of $m$.
The parameters are $\mu/m=1.1$, $\phi/m=0.1$ and
$\phi_b/m=0.1$.}

\label{fig-all}
\end{figure}

The dispersion 
relations for 10 massive particles
are quite complicated. Therefore we performed numerical
calculations 
to extract the corresponding dispersion relations. They are shown 
in Fig.~\ref{fig-all}. 
Bold and thin lines correspond 
to double degenerate and nondegenerate modes, respectively.
As one can see, the two branches connected with gapless NG
modes, contain a roton like excitation at $k_3 \sim m$. Other
characteristic features of the spectrum are also similar
to those of the spectrum for the case with $g^{\prime}=0$
shown in Fig. \ref{fig-all-0}.

\section{Summary}

It would be appropriate to indicate the connection of
our results with related results in the literature.
The possibility of a condensation of vector bosons in
electroweak theory in the presence of a superdense fermionic
matter was considered in Ref. \cite{linde}. This scenario,
with some variations, was further studied in Ref. \cite{shabad}.
A possibility of a vector condensation in two-color QCD
with a baryon chemical potential was suggested in Ref.
\cite{lenaghan}.
Recently, the possibility of a condensation of vector bosons
has been studied in a model at finite density that includes
only massive vector bosons,
with no scalars and fermions \cite{sannino}. The
model is nonrenormalizable and
the authors allow
independent (i.e., not constrained by gauge invariance)
triple and quartic coupling constants.
A sample of spontaneous
symmetry breaking in that model is very different
from that obtained in the present paper.

In conclusion, we studied dynamics in gauged $\sigma$-model at finite
density. For positive $m^2$, the spontaneous breakdown of
$SU(2) \times U(1)_{Y}$ symmetry, caused by a supercritical
chemical potential for hypercharge, is
always accompanied by spontaneous
breakdown of both rotational
symmetry $SO(3)$ [down to $SO(2)$] and electromagnetic $U(1)_{em}$.
On the other hand, for negative $m^2$, the breakdown of
$SU(2) \times U(1)_{Y}$ is realized in two stages, with
both rotational $SO(3)$ and $U(1)_{em}$ being exact at the
first stage. The realization of
both the NG mechanism and the Anderson-Higgs mechanism
in this model is conventional.

The spectrum of excitations in the model is very rich. In
particular, because of the rotational symmetry breakdown, it
is anisotropic: the dispersion relations with respect to
the longitudinal momentum $k_3$ and the transverse momentum
${\bf k}_\perp $ are very different.
A noticeable point is the existence 
of excitation branches that
behave as phonon like quasiparticles for small momenta and
as roton like ones for large longitudinal momenta. This
suggests that this model can be relevant for anisotropic superfluid
systems.

\begin{acknowledgments}
V.P.G. and V.A.M. are grateful for support from the Natural 
Sciences and Engineering Research Council of Canada.
The work of V.P.G. was supported also by
the SCOPES-projects 7UKPJ062150.00/1 and 7 IP 062607 of the Swiss
NSF.
The work of I.A.S. was supported by Gesellschaft f\"{u}r 
Schwerionenforschung (GSI) and by Bundesministerium f\"{u}r 
Bildung und Forschung (BMBF).
\end{acknowledgments}


\begin{thebibliography}{}

\bibitem{MS} V.~A.~Miransky and I.~A.~Shovkovy,
Phys. Rev. Lett. {\bf 88}, 111601 (2002); hep-ph/0108178.

\bibitem{STV}T.~Sch\"{a}fer, D.~T.~Son, M.~A.~Stephanov, D.~Toublan and
J.~J.~Verbaarschot,
Phys. Lett. B  {\bf 522}, 67 (2001); hep-ph/0108210.

\bibitem{BS} P.~F.~Bedaque and T.~Sch\"{a}fer,
Nucl. Phys. A {\bf 697}, 802 (2002);
D.~B.~Kaplan and S.~Reddy,
Phys. Rev. D {\bf 65}, 054042 (2002).

\bibitem{NC} H.~B.~Nielsen and S.~Chadha, Nucl. Phys.
B {\bf 105}, 445 (1976).

\bibitem{superfluid} E.~M.~Lifshitz and L.~P.~Pitaevskii,
{\it Statistical Physics, Part 2} (Pergamon, New York, 1980);
G.~E.~Volovik, {\it Exotic Properties of Superfluid ${}^{3}$He}
(World Scientific, Singapore, 1992). 

\bibitem{tumbling} 
S.~Raby, S.~Dimopoulos and L.~Susskind,
Nucl.\ Phys.\ B {\bf 169}, 373 (1980);
V.~P.~Gusynin, V.~A.~Miransky and Y.~A.~Sitenko,
Phys.\ Lett.\ B {\bf 123}, 407 (1983).

\bibitem{LL} L.~D.~Landau and E.~M.~Lifshits, {\it Statistical Physics}
(Pergamon, New York, 1980).

\bibitem{Kapusta} J.~I.~Kapusta, Phys. Rev. D {\bf 24},
426 (1981).

\bibitem{HW} H.~E.~Haber and H.~A.~Weldon, Phys. Rev. D {\bf 25},
502 (1982).

\bibitem{CW} S.~Coleman and E.~Weinberg, Phys. Rev.
D {\bf 7}, 1888 (1973).

\bibitem{linde}A.~D.~Linde, Phys. Lett. {\bf86}B, 39 (1979);
I.~Krive, Sov. Phys. JETP {\bf56}, 477 (1982) 
[Zh. Eksp. Teor. Fiz. {\bf83}, 849 (1982)]. 

\bibitem{shabad}E.~J.~Ferrer, V.~de~la~Incera, and A.~E.~Shabad,
 Phys. Lett. B {\bf185}, 407 (1987); 
Nucl.\ Phys.\ B {\bf309}, 120 (1988);
J.~I.~Kapusta, Phys. Rev. D {\bf42}, 919 (1990).

\bibitem{lenaghan} 
J.~T.~Lenaghan, F.~Sannino and K.~Splittorff,
Phys.\ Rev.\ D {\bf 65}, 054002 (2002).


\bibitem{sannino}
F.~Sannino and W.~Sch\"{a}fer,
Phys.\ Lett.\ B {\bf 527}, 142 (2002);
F.~Sannino,
Phys.\ Rev.\ D {\bf 67}, 054006 (2003).

\end{thebibliography}
\end{document}